\documentclass[aps,prb,reprint,preprintnumbers,superscriptaddress,amsmath,amssymb,bibnotes,longbibliography]{revtex4-2}

\usepackage{ulem}
\usepackage{graphicx}
\usepackage{dcolumn}
\usepackage{bm}
\usepackage[colorlinks,linkcolor=blue,anchorcolor=blue,citecolor=blue,urlcolor=blue,filecolor=blue,menucolor=blue,runcolor=blue]{hyperref}

\begin{document}
\title{Inverse melting and intertwined orders in PrCuSb$_2$}
\author{H. Q. Ye}
\affiliation{Center for Correlated Matter and School of Physics,
Zhejiang University, Hangzhou 310058, China}
\author{Y. N. Zhang}
\affiliation{Center for Correlated Matter and School of Physics,
Zhejiang University, Hangzhou 310058, China}
\author{T. Le}
\affiliation{Center for Correlated Matter and School of Physics,
Zhejiang University, Hangzhou 310058, China}
\author{H. Q. Yuan}
\email{hqyuan@zju.edu.cn}
\affiliation{Center for Correlated Matter and School of Physics,
Zhejiang University, Hangzhou 310058, China}
\affiliation  {State Key Laboratory of Silicon and Advanced Semiconductor Materials, Zhejiang University, Hangzhou 310058, China}
\affiliation  {Collaborative Innovation Center of Advanced Microstructures, Nanjing 210093, China}
\author{M. Smidman}
\email{msmidman@zju.edu.cn}
\affiliation{Center for Correlated Matter and School of Physics,
Zhejiang University, Hangzhou 310058, China}
\date{\today}
\addcontentsline{toc}{chapter}{Abstract}

\begin{abstract}

Much of the rich physics of correlated systems is manifested in the diverse range of intertwined ordered phases and other quantum states that are associated with different electronic and structural degrees of freedom. Here we find that PrCuSb$_2$ exhibits such phenomena, which at ambient pressure exhibits a fragile antiferromagnetic order, where  cooling in a small $c$ axis magnetic field leads to an additional transition to a field-induced ferromagnetic state. This corresponds to an ‘inverse melting’ effect, whereby further cooling the system restores symmetries of the paramagnetic state broken at the antiferromagnetic transition. Moreover, hydrostatic pressure induces an additional first-order transition at low temperatures, which despite being not likely associated with solely magnetic degrees of freedom, is closely entwined with the magnetic order, disappearing once antiferromagnetism is destroyed by pressure or magnetic fields. Consequently, PrCuSb$_2$ presents a distinct scenario for interplay between different orders, underscoring the breadth of such behaviors within one family of correlated materials.

\end{abstract}

\maketitle

\section{INTRODUCTION}

The coexistence and interplay of different  phases associated with different degrees of freedom is a hallmark feature of strongly correlated electron systems. Archetypal examples are the cuprate superconductors, where unconventional superconductivity manifests in close proximity to antiferromagnetism, together with behaviors such as charge density wave order, a pseudogap phase and strange metals \cite{proust2019remarkable,chang2012direct,Daou2009,badoux2016change}. Superconductivity in Fe-based superconductors similarly arises near antiferromagnetic (AFM) order, and in some cases there is an electronic nematic order which breaks the rotational symmetry of the lattice, that may be driven by spin or orbital ordering \cite{fernandes2014drives,Chu2010,avci2014magnetically,baek2015orbital}. $f$-electron systems also exhibit a range of such phenomena, which often arise from competing interactions, such as between magnetic exchange interactions and the Kondo effect \cite{RevModPhys.81.1551,RevModPhys.73.797,weng2016multiple}. For instance, the heavy fermion superconductor CeCoIn$_5$ hosts a field-induced spin-density wave state (Q-phase) that is strongly intertwined with a spatially modulated superconducting order parameter, vanishing together with superconductivity at the upper critical field \cite{kenzelmann2008coupled,PhysRevLett.104.127001,Gerber2014,PhysRevX.6.041059}. Meanwhile,  applying high magnetic fields to CeRhIn$_5$ induces a Fermi surface reconstruction that at some field angles is concomitant with a nematic state breaking the four-fold lattice symmetry \cite{jiao2015fermi,Moll2015,ronning2017electronic}. Correlated intermetallics can also exhibit an intricate interplay between ferromagnetic (FM) and antiferromagnetic (AFM) orders, whereby applying pressure to suppress a FM transition can induce an AFM or spin-density wave ground state rather than a FM quantum critical point \cite{Belitz1997,brando2016metallic}, as realized experimentally in both transition-metal \cite{Taufour2016,Friedemann2018} and $f$-electron \cite{Sidorov2003,Kotegawa2013,Lengyel2015} magnetic systems.

\begin{figure}[tb]
\includegraphics[width=1\columnwidth]{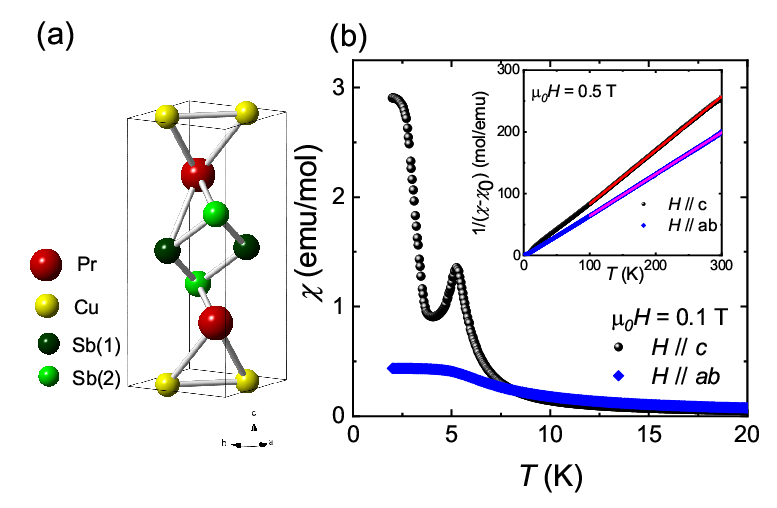}
\caption{(Color online)  (a) Crystal structure of PrCuSb$_{2}$ where the red, yellow and green atoms correspond to  Pr, Cu and Sb, respectively. (b) Low temperature magnetic susceptibility $\chi(T)$ of PrCuSb$_{2}$ with a magnetic field  $\mu_0H=0.1$~T applied parallel to the $c$ axis and within the $ab$ plane. The inset displays the inverse magnetic susceptibility up to 300~K in a 0.5~T field, where the solid lines display the results from fitting with Curie-Weiss behavior.
 }
\label{fig1}
\end{figure}

The $RT$(Sb,Bi)$_2$ ($R$ = Ce, Pr; $T$ = transition metal) series also have a layered tetragonal structure [Fig.~\ref{fig1}(a)],  where the rare-earth ions exhibit a rich range of magnetic behaviors which interplay with the crystalline-electric field (CEF) and Kondo lattice effects \cite{sologub1994crystal,myers1999systematic,thamizhavel2003low,PhysRevB.68.054427,Jobiliong2005,PhysRevB.90.235120}. Among these, it was found that the AFM order of CeAuSb$_2$ below $T\rm_N=6.3$~K is closely preceded by a nematic transition at $T\rm_{nem}=6.5$~K that breaks the $C_4$ lattice symmetry \cite{PhysRevX.10.011035}. Correspondingly, neutron diffraction reveals single-\textbf{q} stripe antiferromagnetism below $T\rm_N$ which also lacks the  rotational symmetry of the tetragonal lattice, while applying a $c$~axis field induces a metamagnetic transition to a double-\textbf{q} structure which restores   $C_4$ symmetry \cite{PhysRevLett.120.097201}. Strong coupling of the order parameters associated with structural and magnetic degrees of freedom is further revealed from pressure dependent measurements, where both $T\rm_N$  and  $T\rm_{nem}$ are gradually suppressed by pressure, reaching a multicritical point which is proposed  to connect the high temperature paramagnetic and  nematic phases, as well as the stripe and multi-\textbf{q} AFM phases \cite{PhysRevX.10.011035}. Moreover, uniaxial strain induces a new spin-density wave (SDW) phase with a distinct propagation vector \cite{Park2018,Waite2022}, suggesting coupling between itinerant SDW orders and a Fermi surface reconstruction.

Here, we find that a Pr-based member of the same series, PrCuSb$_2$, exhibits a very different scenario for intertwined orders. PrCuSb$_2$ was suggested to be ferromagnetic  based  solely on magnetization measurements of polycrystalline samples \cite{sologub1994crystal}, while magnetic Bragg peaks were not resolved in powder neutron diffraction \cite{kolenda2001magnetic}. Our measurements of single crystals demonstrate that  PrCuSb$_2$ exhibits a fragile antiferromagnetism in zero-field with a N\'{e}el temperature $T_{\rm N}=5.2$~K, where a very small  field applied along the $c$~axis induces a metamagnetic transition to a field-induced FM state. The field at which this metamagnetic transition occurs increases with temperature, and therefore the AFM order exhibits an `inverse melting' effect below a transition $T_{\rm M}$ upon cooling in small easy axis fields. Moreover, a moderate pressure induces an additional first-order transition $T_0$ below $T_{\rm N}$ that is not likely solely magnetic in nature, yet is strongly intertwined with the magnetic order, since it vanishes once the magnetic transitions disappear in applied magnetic  fields or pressure. Above  13~GPa, no transitions are detected, and the resulting temperature-pressure phase diagram poses the question as to whether the AFM order disappears below $T_0$, even in the absence of an applied field.

\section{METHODS}

Single crystals of PrCuSb$_{2}$ were grown using a Sb self-flux method with a molar ratio of Pr:Cu:Sb of 1:2:19 \cite{myers1999systematic}. Starting materials of Pr ingots (99.9$\%$), Cu slugs (99.999$\%$) and Sb slugs (99.99$\%$) were loaded into an alumina crucible which was sealed in an evacuated quartz tube. The tube was heated to 1000$~^\circ$C and held at this temperature for two days, cooled slowly to 670$~^\circ$C at 2.75 K/hour, after which  the tube was removed from the furnace and centrifuged to remove excess Sb. The obtained crystals are platelike with typical dimensions $4\times4\times0.3$~mm$^3$. Single crystals of the non-magnetic analog LaCuSb$_2$ were also obtained using a similar procedure. The phase was checked by measuring x-ray diffraction using a XPert MRD (CuK$\alpha$) powder diffractometer  with Cu-K$\alpha$ radiation, where all the Bragg peaks are well-indexed by the (00$l$) reflections of PrCuSb$_2$ \cite{sologub1994crystal}, demonstrating that the $c$ axis is perpendicular to the large face of the crystals. Resistivity and specific heat measurements were performed in  applied fields up to 14~T using a Quantum Design Physical Property Measurement System (PPMS-14) down to 1.8~K, and to 0.3~K using a $^3$He insert. Magnetization measurements were performed in the range 1.8 - 300 K in applied fields up to 5~T using a Quantum Design Magnetic Property Measurement System (MPMS) SQUID magnetometer. Resistivity measurements under pressure were carried out in a piston cylinder cell, and a BeCu diamond anvil cell~(DAC) with a 400-$\mu$m-diameter culet and a Re gasket. Daphne oil 7373 was used as the pressure-transmitting medium.

\section{results}

\subsection{Inverse melting in PrCuSb$_2$}

\begin{figure}[t]
\includegraphics[width=0.8\columnwidth]{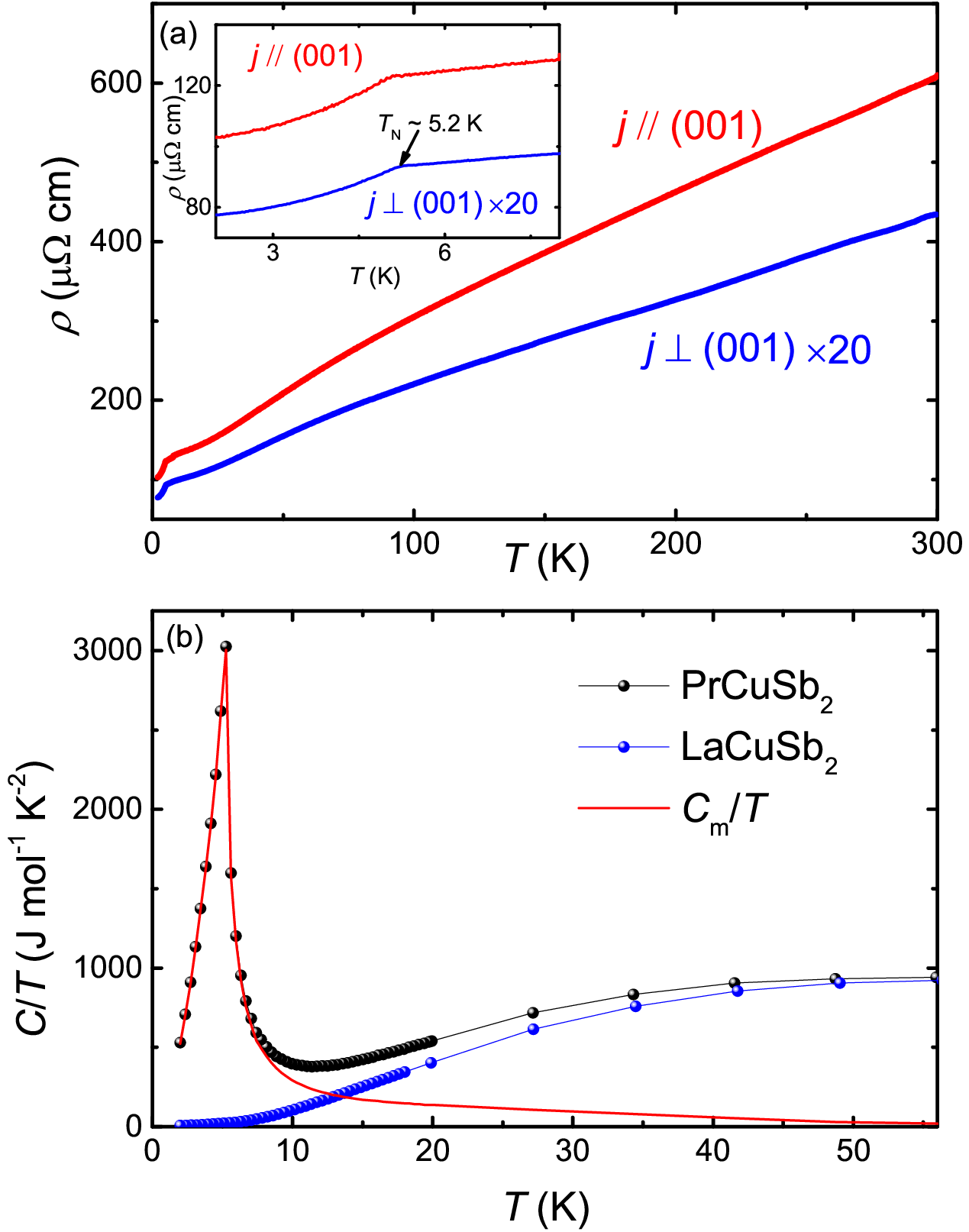}
\caption{(Color online) (a) Temperature dependence of the resistivity $\rho(T)$ of PrCuSb$_{2}$ between 2 and 300~K with the current parallel   (red)  and perpendicular (blue) to the $c$ axis. For clarity, the data for the current perpendicular to the $c$ axis are scaled by a factor of 20. The inset displays the low temperature resistivity, where there is a sharp anomaly at the antiferromagnetic transition. (b) Temperature dependence of the  specific heat (as $C/T$) of PrCuSb$_{2}$ as well as the  nonmagnetic analog LaCuSb$_{2}$, measured down to 2~K. The red solid line shows the magnetic contribution $C\rm_m/T$.
}
\label{fig2}
\end{figure}

\begin{figure}[t]
\includegraphics[width=0.8\columnwidth]{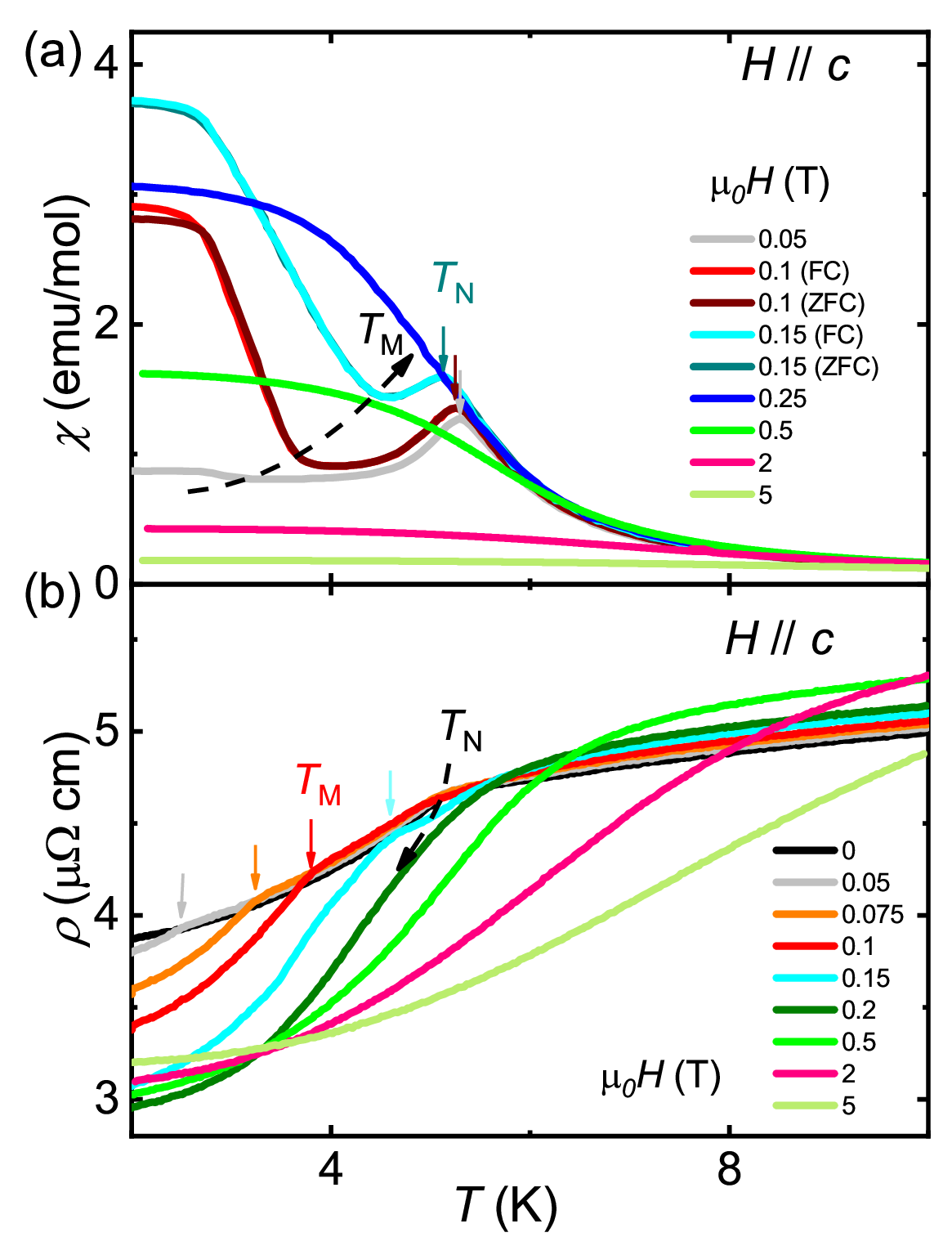}
\caption{(Color online) (a) Low temperature $\chi(T)$ for different magnetic fields applied parallel to the $c$ axis. Measurements for all fields were performed upon warming from the base temperature after zero-field cooling (ZFC), while for 0.1 and 0.15~T both  ZFC and field-cooling (FC) data are displayed. (b) Low temperature $\rho(T)$ of PrCuSb$_{2}$ in various applied magnetic fields parallel to the $c$ axis. The dashed arrow shows the evolution of the AFM transition at $T_{\rm N}$, while the solid arrows highlight the field induced transition $T_{\rm M}$. 
}
\label{fig3}
\end{figure}

\begin{figure}[t]
\includegraphics[width=0.9\columnwidth]{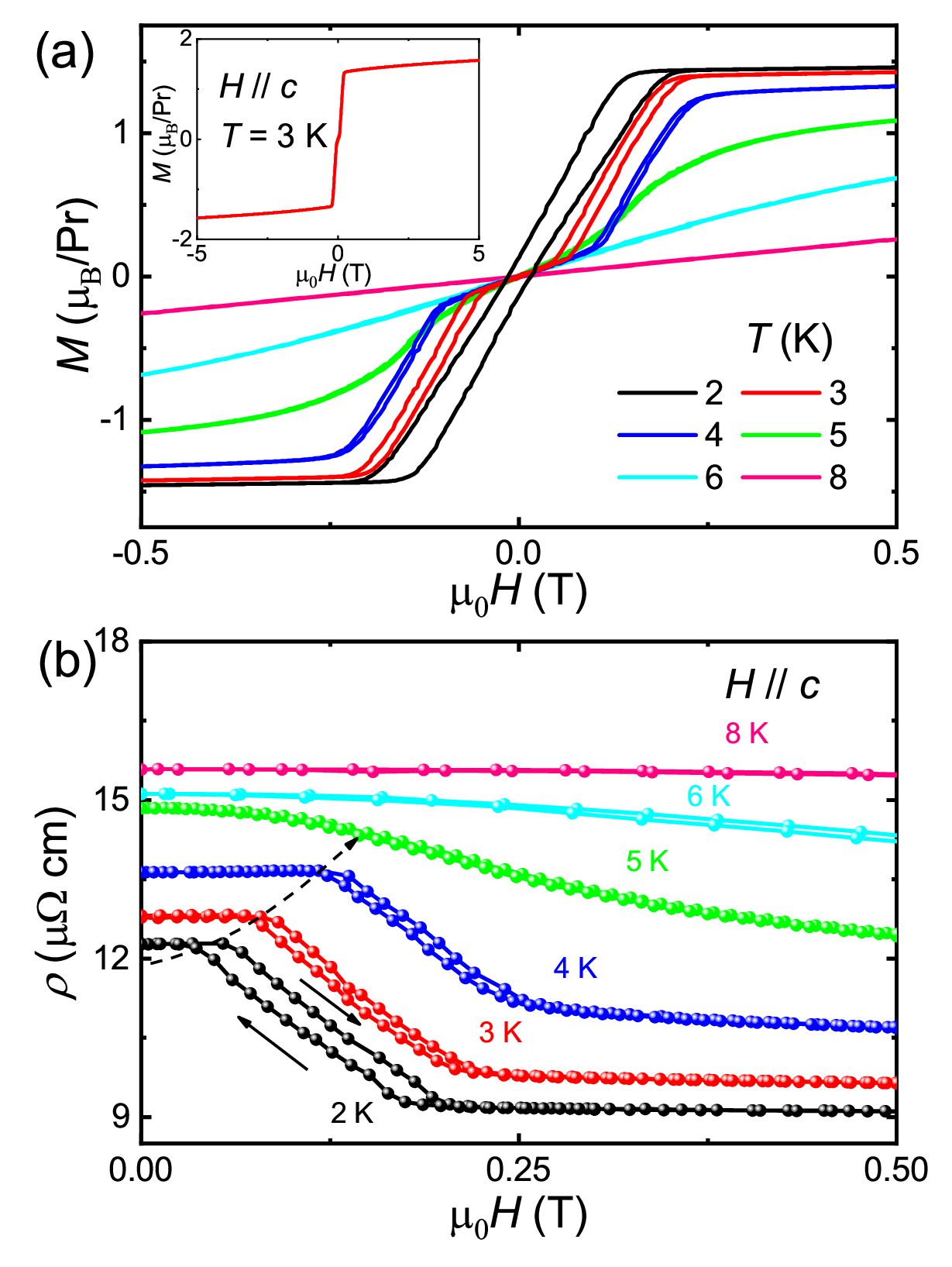}
\caption{(Color online) Field dependence of the (a) magnetization $M(H)$,  and (b) resistivity $\rho(H)$, of PrCuSb$_2$ for fields along the $c$ axis at several temperatures below and above $T_{\rm N}$. The inset of (a) shows $M(H)$ at 3~K in applied fields up to 5~T along the $c$ axis. The dashed arrow highlights the evolution of the metamagnetic transition with temperature, and the solid arrows show the direction of the field sweep. 
}
\label{fig4}
\end{figure}

\begin{figure}[h]
\includegraphics[width=\columnwidth]{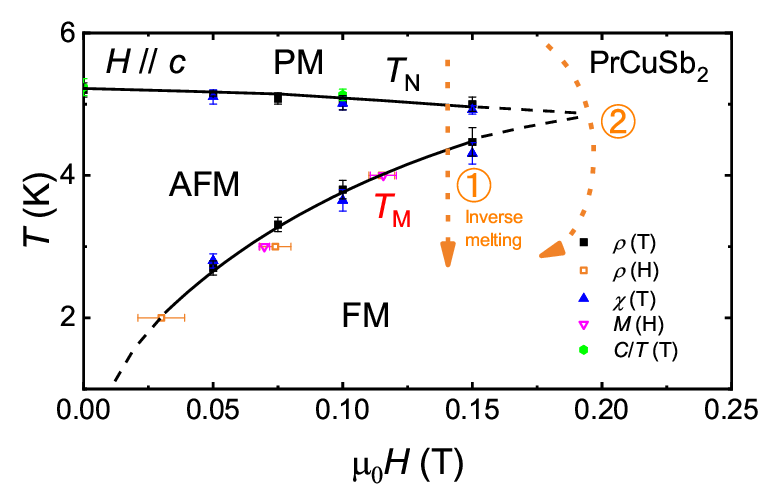}
\caption{(Color online) Temperature-field phase diagram of PrCuSb$_2$ at ambient pressure for fields along the easy $c$ axis, from measurements of the resistivity, susceptibility, magnetization, and specific heat, where the black lines are guides to the eye for $T_{\rm N}$ and $T_{\rm M}$. The orange dotted lines illustrate two paths in the phase diagram, described in the text. Path \textcircled{1} demonstrates an inverse melting effect in PrCuSb$_2$, whereby there is AFM order below $T_{\rm N}$, which melts upon further cooling through  $T_{\rm M}$. On the other hand, path \textcircled{2} demonstrates how the high temperature PM state and low temperature FM state can be connected without going through a phase transition (similar to the liquid-gas phases), and hence these two phases have the same symmetry.}
\label{fig5}
\end{figure}

The low temperature magnetic susceptibility $\chi(T)$ of a PrCuSb$_{2}$  single crystal in a 0.1~T applied field is displayed in Fig.~\ref{fig1}(b), which evidences a magnetic transition  at $T_{\rm N}=5.2$~K. For $H\parallel c$, $\chi(T)$ has a peak at $T_{\rm N}$, while $\chi(T)$ for  $H\parallel ab$ remains almost constant below the transition. Together with the magnitude of $\chi(T)$ at low temperatures being larger for $H\parallel c$ than for $H\parallel ab$, this suggests AFM ordering below $T_{\rm N}$ with the moments orientated along the easy $c$ axis. Furthermore, below the peak at $T_{\rm N}$,  $\chi(T)$ for $H\parallel c$ shows a sizeable increase with decreasing temperature, pointing to the onset of a FM component to the magnetism in a 0.1~T field. The inverse susceptibility measured in a 0.5~T field is displayed in the inset up to 300~K, which above 100 K were analyzed using the Curie-Weiss law: $\chi$=$\chi_0$+$C$/$(T-\theta_{\rm CW})$, where $\chi_0$ is a temperature-independent term, $C$ is the Curie constant and $\theta_{\rm CW}$ is the Curie-Weiss temperature. The fitted results are shown by the solid lines, yielding $\theta^c_{\rm CW}=4.49$~K with an effective moment of  $\mu\rm_{eff}^c$ = 3.03$\mu_B$/Pr  for $H\parallel c$, while $\theta^{ab}_{\rm CW}=5.21$~K and $\mu\rm_{eff}^{ab} = 3.44\mu_B$/Pr for $H \parallel ab$. The obtained values of $\mu\rm_{eff}$ for both directions are close to the full value of  $3.58~\mu_B$ for the $J=4$ ground state multiplet of Pr$^{3+}$. Note that at elevated temperatures  $\chi(T)$ is larger for $H\parallel ab$ than for $H\parallel c$, pointing to a crossover of the easy axis direction.

The temperature dependence of the resistivity  $\rho(T)$  in zero-field also exhibits an anomaly corresponding to the magnetic transition at $T_{\rm N}=5.2$~K [Fig.~\ref{fig2}(a)], where there is an abrupt change of slope. In the paramagnetic (PM) state, $\rho(T)$ is metallic for both current directions, but the overall magnitude is considerably smaller for $j\perp c$ than $j\parallel c$, where  the latter is about 30 times larger at 300~K. This could reflect a quasi-two-dimensional nature of the electronic state, where similar anisotropies were observed in CeAgBi$_2$\cite{thamizhavel2003low} and CeCuSb$_{2}$\cite{PhysRevB.68.054427}.

The transition at $T_{\rm N}$ is also observed in the heat capacity of PrCuSb$_2$ shown in Fig.~\ref{fig2}(b), while the transition is absent in  nonmagnetic isostructural  LaCuSb$_2$. The magnetic contribution to the specific heat coefficient $C{\rm_m}/T$ of PrCuSb$_{2}$  is shown by the red solid line (Appendix), and it can be seen that there is a sizeable contribution above $T_{\rm N}$, which forms an extended tail that reaches up to around 60~K. At $T_{\mathrm{N}}$, the magnetic entropy $S_{\mathrm{m}}$ (Appendix Fig.~\ref{fig9}) reaches $0.93R\ln 2$, which is close to the full value for a ground state doublet. Meanwhile $S_{\mathrm{m}}$ reaches $2R\ln 2$ at 60~K, suggesting that  the aforementioned tail in $C{\rm_m}/T$ is associated with a low-lying excited CEF (doublet) level.

Measurements of $\chi(T)$ with different fields applied along the $c$ axis are shown in Fig.~\ref{fig3}(a). For applied fields less than 0.25~T along the $c$ axis, the peak in $\chi(T)$ corresponding to $T_{\rm N}$ decreases slightly with field. While upon cooling to lower temperatures, there is a pronounced increase of $\chi(T)$ consistent with the presence of an additional transition in applied fields (labelled $T\rm_{M}$), and its position  is determined  from the temperature below which this increase  occurs.  This anomaly is more prominent in  fields of 0.1~T and 0.15~T, and such an increase points to a ferromagnetic nature of the field-induced phase. Above 0.25~T, there are no clear transitions, and there is a broad shoulder which tends towards saturation at the lowest temperatures. The lack of a transition at 0.25~T is also supported by heat capacity measurements  [Appendix Fig.~\ref{fig9}(b)], which exhibit a broad hump instead of the sharp transition observed at lower fields.

The evolution of the AFM and field-induced transitions can be inferred from $\rho(T)$ at low temperatures displayed in Fig.~\ref{fig3}(b), where  $T_{\rm N}$  is slightly suppressed with increasing field. Moreover, while only a single transition is observed in zero-field, in an applied field of 0.05~T, there is an additional anomaly at a lower temperature of 2.5~K,  marked by a vertical arrow, indicating the presence of a field-induced magnetic transition. This field-induced transition $T\rm_{M}$, defined as where $\rho(T)$ exhibits a second abrupt slope change upon cooling, moves to higher temperature with increasing field, as expected for a FM transition. It can be seen that $T\rm_{M}$ approaches $T\rm_{N}$, and at 0.2~T no abrupt low temperature anomalies are observed, and instead  there is a broad hump in $\rho(T)$, which shifts to higher temperature with increasing field.

The isothermal magnetization as a function of field $M(H)$ is displayed in Fig.~\ref{fig4}(a) for fields along the $c$ axis, where the sample was cooled in zero-field, after which a field was applied, and then magnetization loops were measured with both positive and negative fields. The 3~K data exhibits distinct behavior for the two magnetic phases, where  there is no remanent magnetization or hysteresis at zero-field, and $M(H)$ increases linearly at low fields.  At higher fields there is a metamagnetic transition (at $0.067$~T for the up field sweep), at which there is a significant increase of $M(H)$ with hysteresis between up and down field sweeps. Above around $0.23$~T, $M(H)$ changes little with field, suggesting that this corresponds to a saturation magnetization of around 1.4$\mu_{\rm B}$/Pr. As shown in the inset, no further metamagnetic transitions are observed up to at least 5~T. Upon increasing the temperature below $T_{\rm N}$, the low field linear region widens and the hysteresis loops narrow, while at 6~K ($T>T_{\rm N}$) $M(H)$ smoothly increases with field. These behaviors are consistent with an AFM state in PrCuSb$_2$ at zero and very low field, while a small applied field induces a metamagnetic transition to the field-induced FM state. Note that at 2~K the field-induced FM state appears to remain metastable even in zero-field, but when the magnetization is measured with increasing field after zero-field cooling, the
low field metamagnetic transition is still observed [Appendix Fig.~\ref{fig11}(b) inset]. On the other hand, no metamagnetic transitions are observed in either $\rho(H)$ or $M(H)$ for fields applied in the $ab$-plane (see Appendix Fig.~\ref{fig11}).

Metamagnetic transitions are also revealed in the field dependence of the resistivity $\rho(H)$ for fields along the $c$ axis, as displayed in Fig.~\ref{fig4}(b). At 2~K, the resistivity remains nearly constant at very low fields in the AFM phase. At around 0.05~T, there is a metamagnetic transition where there is a significant drop of  $\rho(H)$, with clear hysteresis between measurements performed while sweeping the field up and down, which corresponds to entering the field-induced phase. Upon increasing the temperature below $T_{\rm N}$, the metamagnetic transition moves to higher fields, while at 5~K just below $T_{\rm N}$, there is a broad decrease of $\rho(H)$ with weak hysteresis.

The temperature-field phase diagram of PrCuSb$_2$ for $c$~axis fields based on the above measurements is displayed in Fig.~\ref{fig5}, where the phase boundaries deduced from different techniques are highly consistent.  In zero field there is a single AFM transition below $T_{\rm N}=5.2$~K. Upon applying a small $c$ axis field, $T_{\rm N}$ is suppressed slightly, but a field-induced transition $T_{\rm M}$ to a FM state occurs below $T_{\rm N}$, which increases rapidly with field. Above around 0.2~T, no transitions are resolved, and instead there is a crossover to the spin-polarized FM phase. It can be seen that if the system is cooled in a moderate easy axis field (following the path labelled \textcircled{1} in the phase diagram), the system first undergoes a transition from the PM state to the AFM phase, and at lower temperatures there is a second transition from the AFM to field-induced FM state. It is possible to reach the field-induced FM state from the high temperature paramagnetic (PM) phase without going through a phase transition (path \textcircled{2}), demonstrating in analogy with the liquid-gas transition that the two phases have the same symmetry (time-reversal symmetry is broken in the PM state by the applied field). Conversely, translational and/or rotational symmetries of the PM state will be broken entering the AFM phase. As such, the transition from the AFM to FM phase corresponds to an `inverse melting' effect, whereby upon cooling below  $T\rm_{M}$ there is a transition from the lower symmetry AFM phase, to one with higher symmetry \cite{Schupper2004,Paoluzzi2010,Jacksonthesis}. We note that if the transition $T\rm_{M}$ corresponded to a transition to a different type of magnetic ordering rather than an FM phase, a phase transition would be expected between this phase and the high temperature PM region, which is not detected. The spin-polarized nature of this field-induced phase could be verified by measurements such as neutron diffraction.

\begin{figure}[t]
\includegraphics[width=0.95\columnwidth]{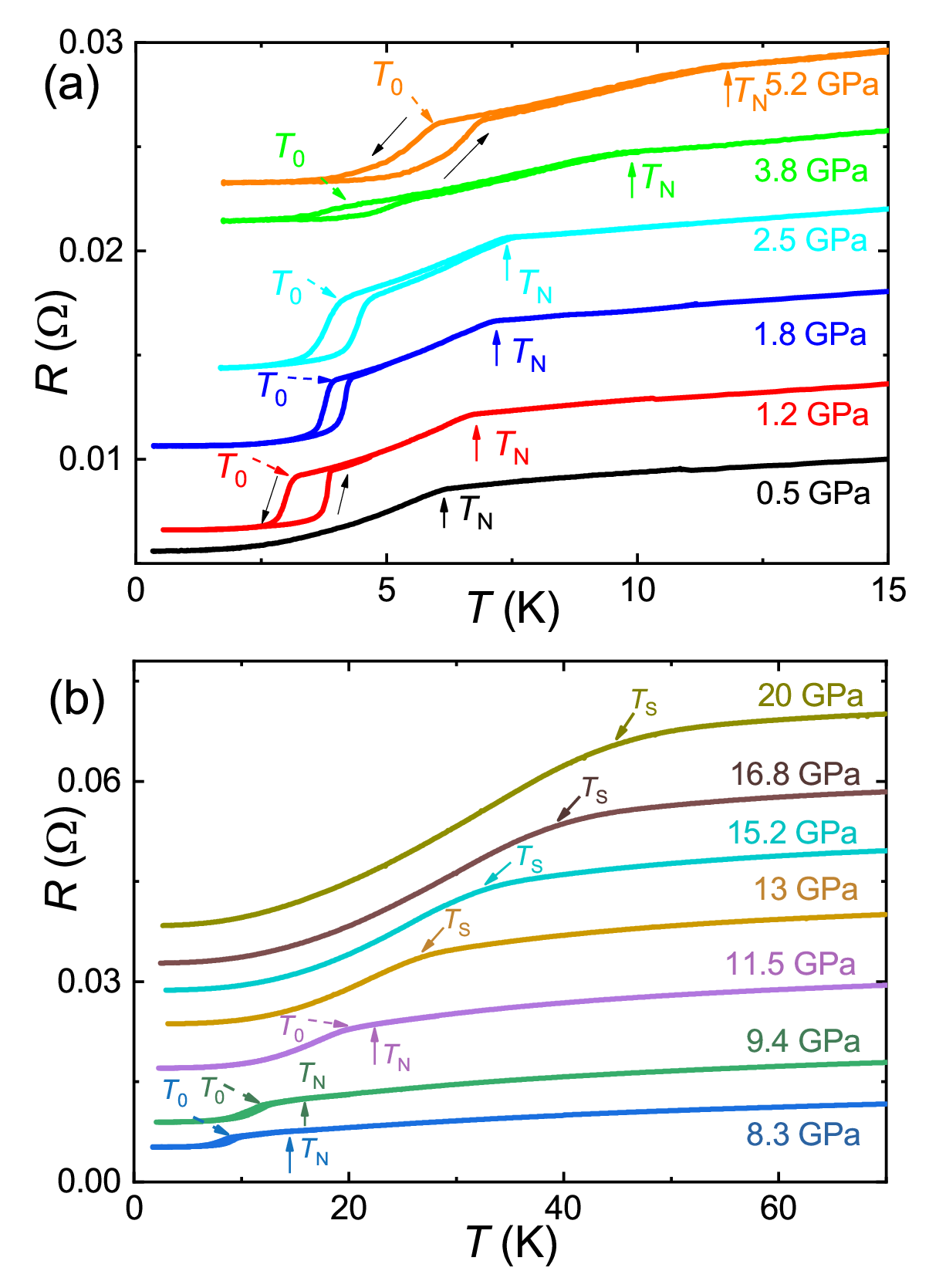}
\caption{(Color online) Temperature dependence of the resistance $R(T)$ of PrCuSb$_2$ in zero applied field (a) at low pressures up to 5.2~GPa,  and (b) at higher pressures up to 20~GPa. For clarity, the data for adjacent pressures have been shifted vertically.  The arrows denote the antiferromagnetic transition at $T\rm_{N}$, the pressure induced first-order transition at $T_0$, and the broad shoulder feature $T\rm_{S}$. The shoulder feature at high pressures is characterized by an increase of the derivative of $R(T)$ over a broad temperature range upon cooling, and  $T\rm_{S}$ is defined as the midpoint of this increase. Note the measurements at  2.5~GPa and below were performed in  a piston cylinder cell, while those at higher pressures used  a diamond anvil cell.
}
\label{fig6}
\end{figure}

\subsection{Intertwined phases under pressure }

To determine the evolution of the magnetic order with pressure, the resistivity of PrCuSb$_{2}$ was measured under different hydrostatic pressures up to 20~GPa, which are displayed for zero applied field in Fig.~\ref{fig6}.
At 0.5~GPa, the data resembles that at ambient pressure, exhibiting a single AFM transition with a slightly enhanced $T\rm_N$. On the other hand, the data at 1.2~GPa are drastically different, where in addition to $T\rm_N$, another anomaly denoted $T_0$ is observed at a lower temperature of around 3.2~K. Upon cooling below $T_0$, the resistive drop  is much more pronounced than at $T\rm_N$, and moreover there is sizeable hysteresis between measurements performed upon sweeping the temperature up and down, pointing to a first-order nature of this pressure-induced transition.  $T\rm_N$ and $T_0$ are enhanced by pressure,  both reaching over 20~K at 11.5~GPa. Since  at higher pressures $T_0$ increases more rapidly than $T\rm_N$, these transitions appear to merge together at around 13~GPa, and at this pressure no pronounced transition is observed, where instead there is a broad shoulder feature in the resistivity. The position of this shoulder is denoted  by $T\rm_{S}$, which increases and broadens upon increasing the pressure up to 20~GPa. These results are summarized by the phase diagram in Fig.~\ref{fig8}(a), and point to a close link between these transitions, which simultaneously disappear upon merging under pressure, being replaced by a broad crossover feature.

\begin{figure}[t]
\includegraphics[width=0.99\columnwidth]{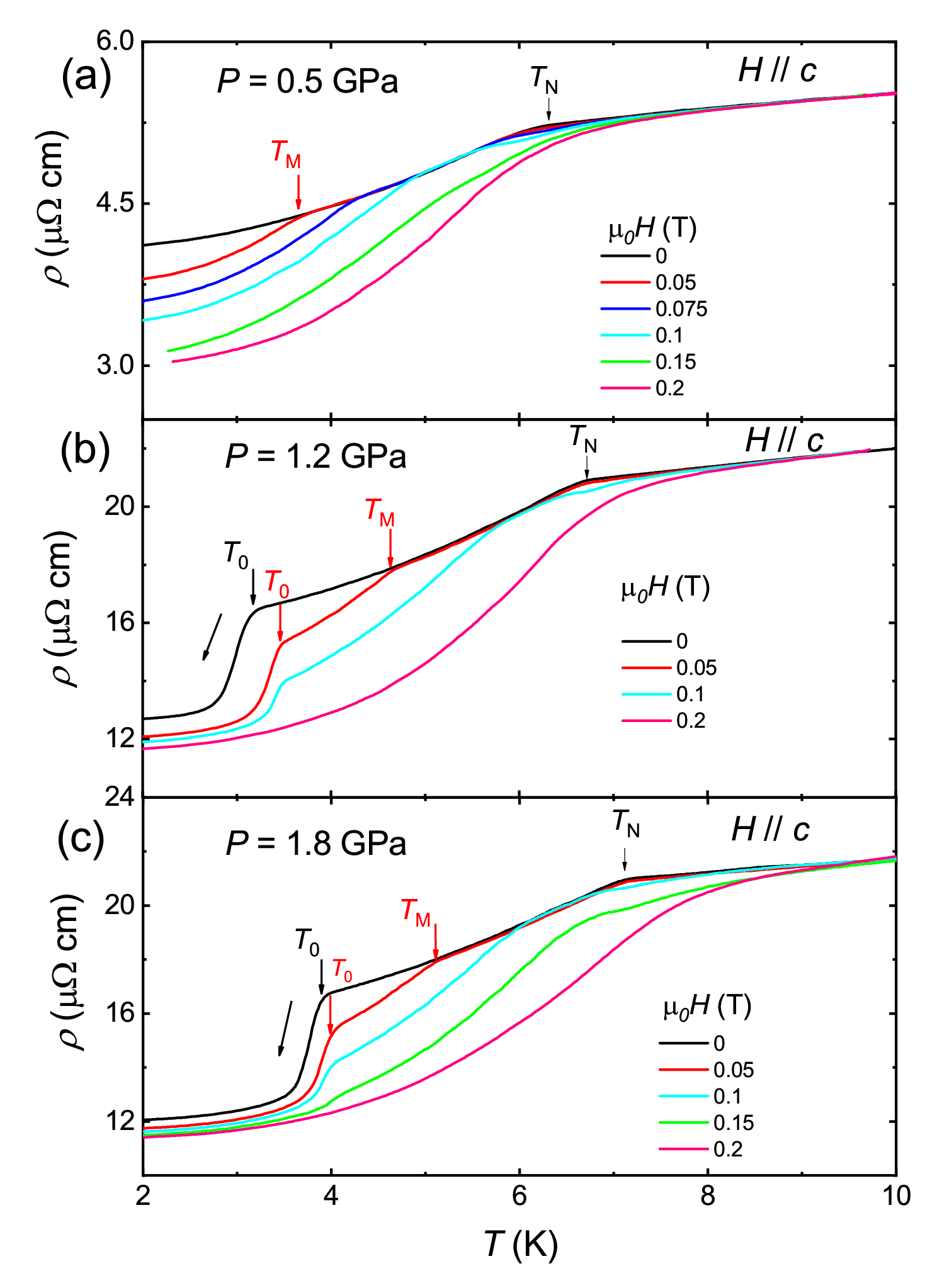}
\caption{(Color online)  $\rho(T)$ of PrCuSb$_2$ under pressure in several applied fields along the $c$ axis at (a) 0.5~GPa, (b) 1.2~GPa and (c) 1.8~GPa. The arrows denote the transitions at $T\rm_N$,  $T_0$, and $T\rm_{M}$. All the measurements were performed upon decreasing the temperature.
 }
\label{fig7}
\end{figure}

\begin{figure}[t]
\includegraphics[width=0.95\columnwidth]{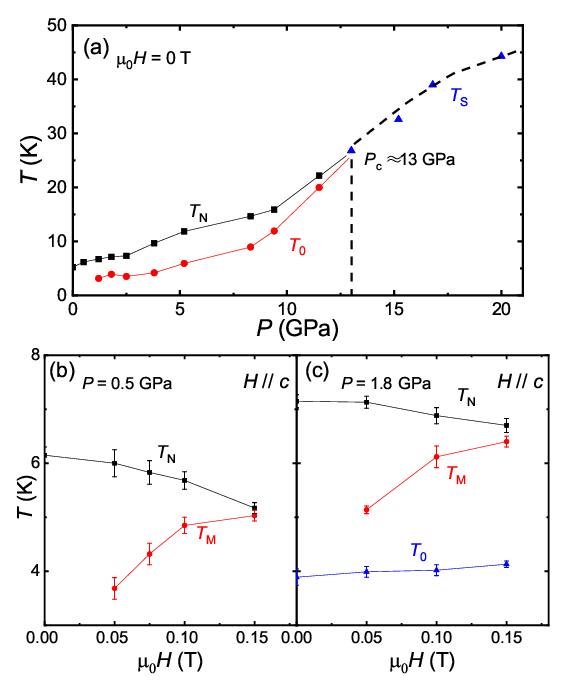}
\caption{(Color online) (a) Temperature-pressure phase diagram of PrCuSb$_2$ at zero field, where the black squares, red circles, and blue triangles  correspond to $T\rm_N$, $T_0$, and $T\rm_{S}$, respectively.  $P\rm_c$ represents the pressure above which $T\rm_N$ and $T_0$ disappear, and there is the emergence of a broad shoulder denoted by $T\rm_{S}$. Temperature-field phase diagrams of PrCuSb$_2$ for fields applied along the $c$ axis at (b) 0.5~GPa and (c) 1.8~GPa.
 }
\label{fig8}
\end{figure}

Upon applying magnetic fields under pressure, $\rho(T)$ at 0.5~GPa  [Fig.~\ref{fig7}(a)] is again very similar to that at ambient pressure [Fig.~\ref{fig3}(b)], where the transitions at $T\rm_N$ and $T\rm_{M}$ are observed in a small applied field, while no transitions are found in 0.2~T. Conversely, upon applying a small 0.05~T field at higher pressures of 1.2  and 1.8 GPa [Figs.~\ref{fig7}(b) and (c)],   three transitions are now observed corresponding to $T\rm_N$, $T\rm_{M}$ and $T_0$, where the latter pressure-induced transition $T_0$ is situated below $T\rm_{M}$. These show that the pressure-induced transition at $T_0$ must be distinct from the field-induced transition at $T\rm_{M}$, and that $T_0$ persists even in the field-induced FM phase. On the other hand, none of the transitions are found in a 0.2~T field, again demonstrating that the disappearance of $T_0$ coincides with that of $T\rm_N$, pointing to a close link between these transitions. These results are summarized by the field-temperature phase diagrams in Figs.~\ref{fig8}(b) and (c). At 0.5~GPa, only $T\rm_N$ and $T\rm_{M}$ can be observed in applied fields, in line with the ambient pressure case. At 1.8~GPa, there are similar behaviors of $T\rm_N$ and $T\rm_{M}$ to those at low pressures, while there is little change of $T_0$ with field, before it abruptly disappears above 0.15~T together with the other transitions.

\section{DISCUSSION AND SUMMARY}

Our measurements of the resistivity, magnetic susceptibility and specific heat at ambient pressure show that PrCuSb$_{2}$ orders antiferromagnetically below $T_{\rm N}=5.2$~K, with the moments orientated along the $c$ axis. A small $c$~axis field induces a metamagnetic transition, above which the spins are polarized with a saturation magnetization of 1.4$\mu\rm_{B}$/Pr. In sufficiently low $c$ axis fields, $\rho(T)$ exhibits a second transition below  $T_{\rm N}$, labelled $T_{\rm M}$, which in comparison to the  measurements of $M(H)$ and $\rho(H)$ [see phase diagram in Fig.~\ref{fig5}] can be seen to correspond to a transition from the AFM to the field-induced FM phase. Since the applied magnetic field already breaks time-reversal symmetry in the high temperature paramagnetic state, this field-induced FM phase does not break any additional symmetries, while in the AFM phase translational and/or rotational symmetries are broken. Consequently this is a manifestation of `inverse melting', where upon cooling through $T_{\rm M}$,  there is a transition to a higher symmetry phase  \cite{Schupper2004,Paoluzzi2010,Jacksonthesis}. It is notable that in the low temperature limit only a very small $c$ axis field is required to polarize the spins ($<0.05$~T), suggesting extremely weak AFM coupling relative to the energy scale of $T_{\rm N}$, indicating the presence of significant FM interactions. Moreover this suggests that in these small fields the field-induced FM state has the lowest energy, while the occurrence of AFM order at higher temperatures indicates that the AFM state is \textit{entropically} favored in a narrow temperature region, pointing to the presence of additional degeneracies associated with the AFM phase \cite{Schupper2004}. Uncovering the mechanism behind this inverse melting effect therefore requires additional probes of the structure and excitations associated with different degrees of freedom, including using techniques such as neutron and x-ray scattering, as well as nuclear magnetic resonance. Note that while   the metamagnetic transition  of  Nd$_{0.5}$Sr$_{0.5}$MnO$_3$ between the charge-ordered insulating AFM phase and a metallic field-induced FM phase can be shifted to higher fields with increasing temperature, this was only realized for measurements performed with decreasing field, and hence the reentrance of the higher symmetry metallic phase is ascribed to the supercooling of a metastable state \cite{Kuwahara1995}. In contrast, the increase of the metamagnetic field with temperature in PrCuSb$_2$ is robust for both increasing and decreasing field-sweeps (Fig.~\ref{fig4}).

Under a moderate pressure of 1.2 GPa, an additional first-order transition $T_0$ emerges in the low temperature resistivity below $T_{\rm N}$. This transition is unlikely to be purely magnetic in nature, since there is a pronounced drop of the resistivity below $T_0$, a far larger decrease than below the magnetic transitions $T_{\rm N}$ and $T_{\rm M}$, together with significant thermal hysteresis. Furthermore, $T_0$ persists in a small applied field, and still exists below $T_{\rm M}$ in the field induced FM state. On the other hand, this transition is  closely coupled to the magnetic order, since in larger applied fields $T_0$ is absent once $T_{\rm N}$ and $T_{\rm M}$ disappear. In addition, both $T_0$ and $T_{\rm N}$ increase with pressure, and these transitions merge at a pressure of around $P_{\rm C}\approx13$~GPa, above which no clear transition is detected, but there is a broad shoulder in the resistivity.

A pronounced resistivity anomaly together with a first-order nature suggests that $T_0$ could correspond to a structural transition that is closely coupled to magnetic order.  In isostructural CeAuSb$_2$,  a structural transition that breaks the $C_4$ lattice symmetry $T_{\rm Nem}$ is observed just above an AFM transition to a single-\textbf{q} striped magnetic phase \cite{PhysRevX.10.011035,PhysRevLett.120.097201}. On the other hand, in PrCuSb$_2$ there is an abrupt disappearance of $T_0$ in field [Fig.~\ref{fig8}(c)] which is suggestive that the first-order $T_0$ line terminates at a finite temperature critical point. In this scenario, $T_0$ would not be anticipated to break any additional lattice symmetries (in analogy with the liquid-gas transition), and hence is not likely to correspond to a nematic transition. Furthermore, given the similarities between the phase diagrams versus field and pressure, it is of particular interest to determine whether there is the disappearance of AFM order below $T_0$. To address this question and the nature of the transition at $T_0$, neutron diffraction under pressure is highly desirable.

In addition, while the magnetic entropy being close to $R$ln2 at $T_{\rm N}$ points to a doublet ground state, there is still a sizeable contribution above $T_{\rm N}$, reaching around 2$R$ln2 at 60~K. This suggests the  presence of low-lying CEF levels, and therefore it would be desirable to determine whether there is also a role played by quadrupolar degrees of freedom \cite{Kosaka1998,Kiss2003,Takimoto2008}. Consequently, measurements such as inelastic neutron scattering to determine the CEF level scheme are important for gaining a microscopic understanding of the different orders.

In conclusion, PrCuSb$_{2}$ exhibits intertwined orders that are readily tuned by magnetic fields and pressure. In zero-field this is manifested by a fragile AFM phase that is readily destroyed by a $c$ axis field, and upon cooling in small applied fields there is `inverse melting' to a low temperature FM state. Meanwhile moderate pressures induce a low temperature first-order transition $T_0$, which while not likely solely associated with magnetic degrees of freedom, is strongly coupled to the magnetism, since it disappears once small fields destroy the magnetic order, and also merges with the AFM transition at high pressures around $P_{\rm C}$ $\approx$ 13~GPa. The intricate interplay of these phases presents differently to  that of the intertwined phases in  CeAuSb$_2$ and the iron-pnictides, in which the coupling between itinerant magnetic (SDW) \cite{Lumsden2010,Waite2022} and nematic orders gives rise to similar phase diagrams \cite{PhysRevX.10.011035,Fernandes2012,fernandes2014drives}. This suggests that despite PrCuSb$_2$ being isostructural to CeAuSb$_2$, it should be described by fundamentally different microscopic and phenomenological models, which may be related to the presence of more localized magnetism. Together, these underscore the rich variety of intertwined orders within a single family of correlated rare-earth intermetallics.

\section*{acknowledgments}

We are grateful to Zhentao Wang for interesting discussions and helpful suggestions. This work was supported by the National Key R$\&$D Program of China (Grants No.~2022YFA1402200 and No.~2023YFA1406303), the Key R$\&$D Program of Zhejiang Province, China (Grant No.~2021C01002), the National Natural Science Foundation of China (Grant No.~12222410, No.~12174332, No.~11974306, and No.~12034017).

\section*{APPENDIX}

\subsection{Specific heat of PrCuSb$_2$}

\begin{figure}[tb]
\includegraphics[width=0.85\columnwidth]{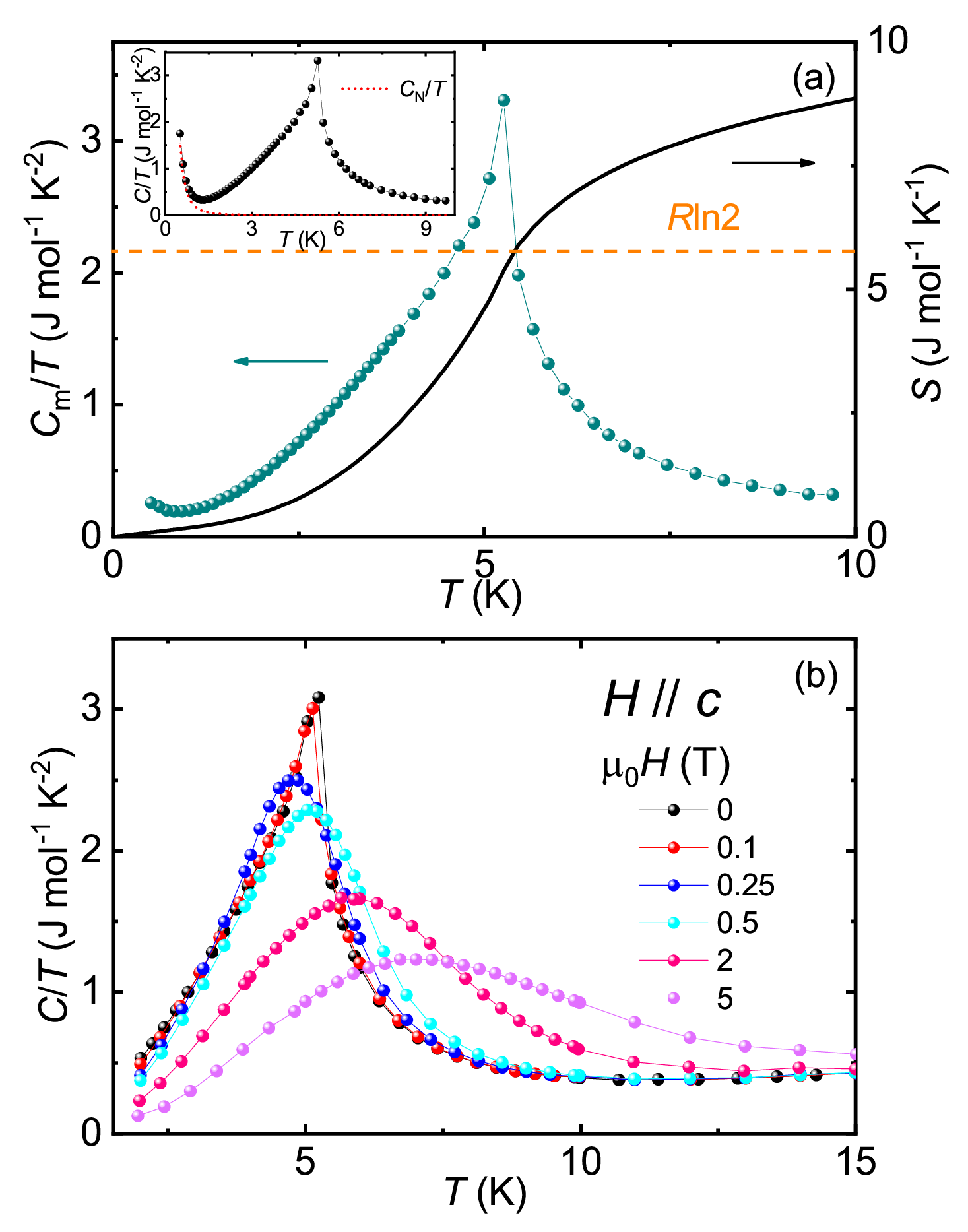}
\caption{(Color online) (a) Low temperature $C{\rm_m}/T$  of PrCuSb$_{2}$ measured down to 0.5~K, together with the  magnetic entropy $S\rm_m$. Here a contribution from a nuclear Schottky anomaly $C{\rm_N}/T$ (red dashed line of the inset) has been subtracted from the estimated Pr contribution $C{\rm_{Pr}}/T$ obtained from subtracting the $C/T$ of  LaCuSb$_{2}$. (b) Temperature dependence of $C/T$ of PrCuSb$_{2}$ in various applied magnetic fields parallel to the $c$ axis.
 }
\label{fig9}
\end{figure}

\begin{figure}[tb]
\includegraphics[width=0.65\columnwidth]{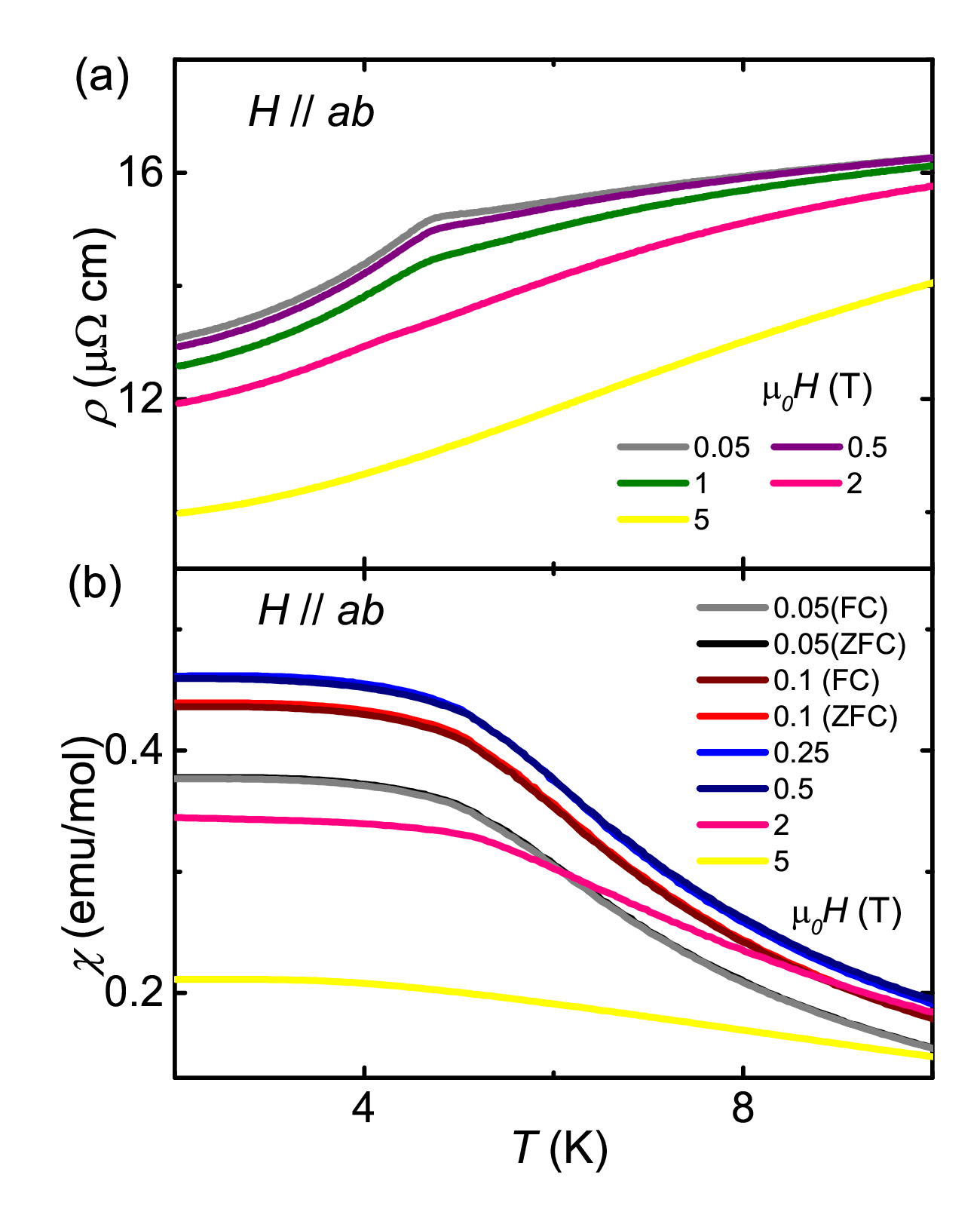}
\caption{(Color online) (a) Low temperature $\rho(T)$ of PrCuSb$_{2}$ in various  magnetic fields applied within the $ab$ plane. (b) $\chi(T)$ for different magnetic fields applied within $ab$ plane. Data are shown for both zero-field cooling (ZFC) and field-cooling (FC), which were performed upon warming from the base temperature.
}
\label{fig10}
\end{figure}

\begin{figure}[tb]
\includegraphics[width=0.7\columnwidth]{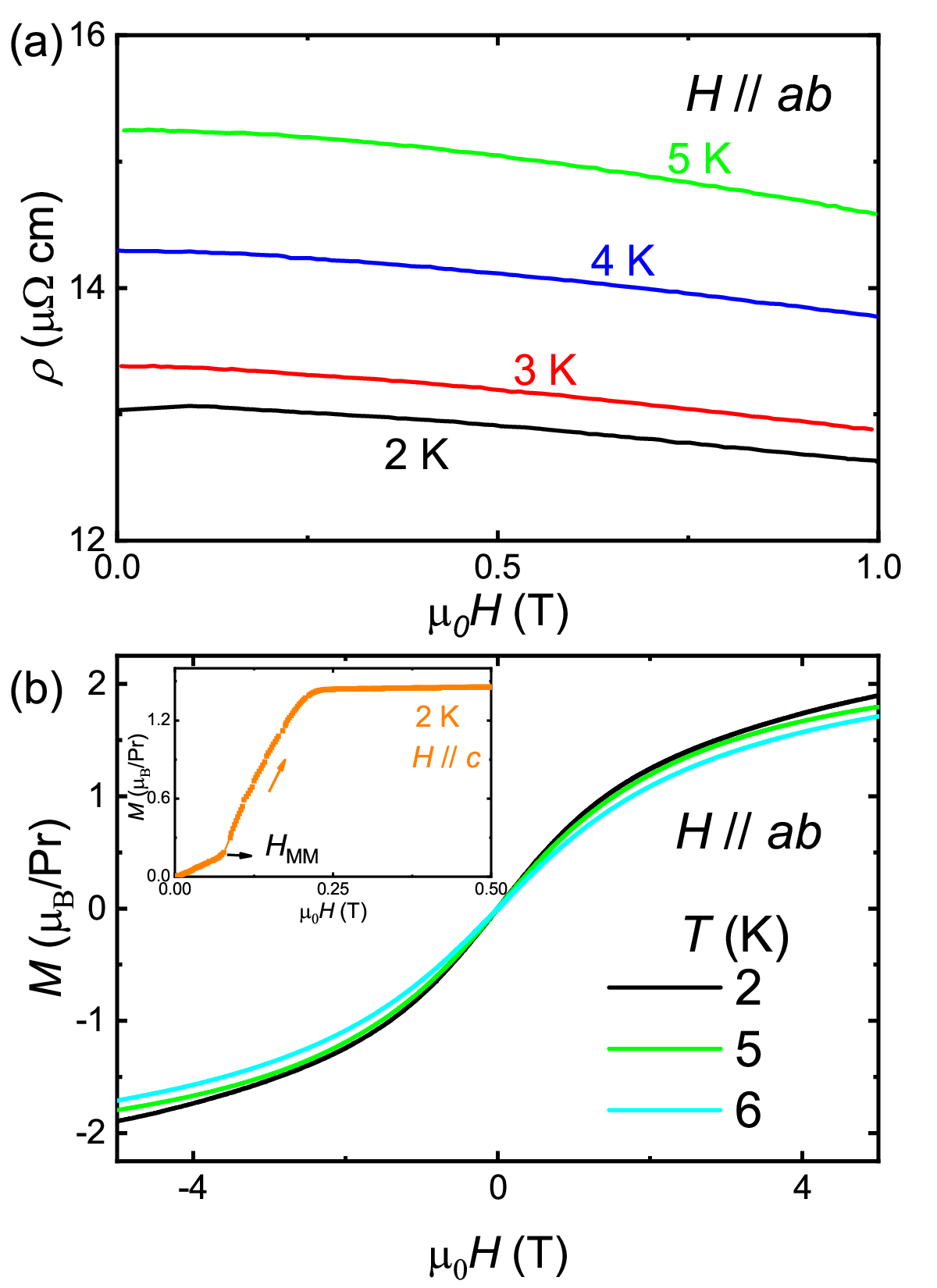}
\caption{(Color online) Field dependence of the (a) resistivity $\rho(H)$,  and (b) magnetization $M(H)$ of PrCuSb$_2$ for fields within the $ab$ plane, at several temperatures below and above $T_{\rm N}$. The inset displays  $M(H)$ of PrCuSb$_2$ for fields along the $c$ axis at 2~K measured with the field increasing after zero-field cooling. It can be seen that the metamagnetic transition is detected at low fields, similar to the 3~K data in Fig.~\ref{fig4}(a).
}
\label{fig11}
\end{figure}

Figure~\ref{fig9}(a) shows the magnetic contribution to the specific heat coefficient $C{\rm_m}/T$ of PrCuSb$_{2}$ down to 0.5~K. As shown in the inset, the low temperature $C/T$ shows a steep upturn arising from a nuclear Schottky anomaly. This low temperature contribution was estimated from fitting a nuclear Schottky anomaly contribution $C\rm_N$ = $A/T^2$ (dashed line). In order to obtain $C{\rm_m}/T$, both the phonon contribution estimated from the data of LaCuSb$_2$ in main text Fig.~\ref{fig2}(b), as well as $C\rm_N/T$, were subtracted. The temperature dependence of the magnetic entropy $S_{\mathrm{m}}$ of PrCuSb$_2$ is also displayed obtained by integrating $C_{\mathrm{m}} / T$. Note that since a tiny residual upturn in  $C_{\mathrm{m}} / T$ is still present from the incomplete subtraction of $C\rm_N$, $C_{\mathrm{m}} / T$ was taken to have a constant value below  0.8~K. At $T_{\mathrm{N}}$, $S_{\mathrm{m}}$ reaches $0.93R\ln 2$, while $S_{\mathrm{m}}$ reaches $2R\ln 2$ at 60~K.  From extrapolating the $C/T$ data for $T>T_{\rm N}$,  a large zero temperature value of $\gamma=215~\mathrm{mJ} / \mathrm{mol}~\mathrm{K}^{2}$ is obtained, which could indicate an enhanced Sommerfeld coefficient arising from strong electronic correlations.

Figure~\ref{fig9}(b) displays the low temperature $C/T$ of PrCuSb$_{2}$ with different fields applied along the $c$ axis. It can be seen that in a field of 0.1~T the transition is slightly suppressed to 5.1~K in line with an AFM transition. Above 0.25~T no pronounced transition is observed, where instead there is  broad peak, which further broadens with increasing field. These suggest that a very small $c$~axis field can tune  PrCuSb$_{2}$ from the AFM to field-induced FM phase.

\subsection{Dependence of the magnetic properties on $ab$ plane fields}

The low temperature $\rho(T)$ for fields within the  $ab$ plane is shown in Fig.~\ref{fig10}(a). Here the AFM transition is more robust than for fields parallel to the $c$ axis [Fig.~\ref{fig3}(b)], and can be detected up to at least 1~T, while above 2~T no clear anomaly is observed. Moreover, no additional field-induced transitions are observed for this applied field orientation.  The low temperature magnetic susceptibility  for fields within the $ab$ plane is shown in Fig.~\ref{fig10}(b), where the position of $T\rm_N$ also changes little with field, and below the transition $\chi(T)$ has a nearly constant value.

The field dependence of the resistivity and magnetization are displayed in Fig.~\ref{fig11}(a) and Fig.~\ref{fig11}(b) respectively, for fields within the $ab$ plane. Neither quantity exhibits any metamagnetic transitions, where $\rho(H)$ gradually decreases with field, while $M(H)$ smoothly increases. In a 5~T field, the magnetization reaches 1.9~$\mu_{\rm B}$/Pr, which is comparable to the corresponding value of 1.6~$\mu_{\rm B}$/Pr for fields along the $c$ axis, suggesting a more isotropic CEF ground state in high fields.

\end{document}